Review

# The role of friction forces in arterial mechanical thrombectomy: a review

Mahesh S. Nagargoje [a,*], Virginia Fregona [a], Giulia Luraghi [a], Francesco Migliavacca [a,b], Guglielmo Pero [c,d], Jose Felix Rodriguez Matas [a]

[a] Computational Biomechanics Laboratory, Laboratory of Biological Structure Mechanics (LaBS), Department of Chemistry, Materials and Chemical Engineering "Giulio Natta", Politecnico di Milano 20133 Milano, Italy
[b] Fondazione IRCCS Cà Granda Ospedale Maggiore Policlinico, Milano, Italy
[c] Department of Medicine and Surgery, Kore University of Enna, Italy
[d] Department of Neuroradiology, AOE Cannizzaro, Catania, Italy

ABSTRACT

Multiple clinical trials have demonstrated the superiority of mechanical thrombectomy (MT) over intravenous thrombolysis (tPA) in treating acute ischemic stroke (AIS). Stent retriever (SR) and aspiration techniques are the standard methods for removing occluded emboli, with evolving technologies improving MT efficiency. However, procedural success remains uncertain. Frictional forces, specifically clot-vessel, clot-SR, and SR-vessel interactions, play a critical role in MT outcomes. This review examines frictional forces during MT and their impact on success, analyzing publications from 2015 to 2025. We focus on studies that calculated friction or retrieval forces using in vitro models. We have also included current trends, limitations, and future perspectives on studying and understanding frictional forces and their implementation into in silico models. Findings indicate that fibrin-rich clots are more difficult to retrieve than red blood cell (RBC)-rich clots due to their higher friction coefficient, three to four times greater, an observation supported by multiple studies. SR-vessel and SR-clot friction also influence MT effectiveness. SR-vessel interaction plays a crucial role in acutely curved vessels, as SR compression reduces its efficiency. In SR-clot interaction, RBC-rich clot fragmentation is linked to relative interaction forces. In summary, obtaining in vivo frictional values remains challenging, and inconsistencies persist in past in vitro studies. Further, a deeper understanding of frictional forces is essential for optimizing MT, improving current SRs, and developing next generation thrombectomy technologies.

## 1. Introduction

Stroke accounts for 6.6 million deaths and it is second leading cause of death globally. Lancet Neurology Commission projects a 50 % rise in stroke mortality by 2050, with annual deaths expected to climb from 6.6 million in 2020 to 9.7 million (Feigin et al., 2023). Acute ischemic stroke (AIS) accounts for approximately 80 % of all stroke cases and poses a significant burden on the world economy due to its management and treatment. Rapid endovascular treatment significantly improves functional outcomes in patients with AIS caused by proximal intracranial arterial occlusions. Early intervention led to higher rates of reperfusion and reduced disability without increasing mortality (Berkhemer et al., 2015; Campbell et al., 2015; Goyal et al., 2015). Despite the dramatic improvement in success rate, there remains an uncertainty and a void for further understanding and improvement. The uncertainty in MT outcomes depends on factors such as the type of SR, type of clot, and SR-vessel interaction (Yoo and Andersson, 2017). SR-tissue-clot interaction causes friction and plays a key role in thrombectomy success. Another factor influencing the thrombectomy outcomes is the pressure gradient across the clot and clot-vessel adhesion, especially during the aspiration procedure (Maïer et al., 2020). Due to increased medical device implantation into human bodies, a new stream has gained importance called bio-tribology. Bio-tribology studies the frictional interactions between medical devices and vessels during medical procedures. A review article by Wagner et al. (2021) discusses stent and catheter interactions and its interaction with vessel tissues, highlighting the limited understanding of tissue tribological mechanism (Wagner et al., 2021). It also critically compares existing experimental approaches and underscores the importance of advancing bio-tribological research to improve device safety and performance. In this review, we have restricted our discussion to mechanical interactions/friction during MT.

* Corresponding author at: Department of Chemistry, Materials, and Chemical Engineering "Giulio Natta', Politecnico di Milano, 20133, Italy.
*E-mail address:* maheshshahadeo.nagargoje@polimi.it (M.S. Nagargoje).

During MT, multiple moving contacts and interactions lead to multiple frictional force combinations. These combinations consist of clot-vessel, clot-SR, and SR-vessel interaction, as shown in Fig. 1. Moreover, we must consider that intracranial vessels are not fixed structures but can move within the subarachnoid space in response to tractions (McCarthy et al., 2022). The static friction between clot and low friction surfaces has been investigated, and it was found that fibrin-rich clots exhibit higher friction compared to red blood cell (RBC)-rich clots (Gunning et al., 2018). Similarly, SR-vessel frictional force is dominant and plays a crucial role during MT (Elkhayyat et al., 2024). SR-vessel interaction is sufficient to induce vasospasm, and intimal damage or pseudoaneurysm formation is exceptional (Singh et al., 2013). In contrast, another study found arterial wall thickening and enhancement in patients treated with MT due to SR-vessel interaction (Power et al., 2014). It has been observed that SRs get compressed in the acutely curved vessels, which might lead to clot dislodgment and failure of thrombectomy (Kaneko et al., 2019; Machi et al., 2017). Clinically, it has been observed that acute curvature leads to bending of the SR, and extravasation has been detected due to higher friction at acute curved region, leading to pseudo-aneurysm formation due to SR compression, which is considered as a rare case (Misaki et al., 2016). It has also been observed that SR removal forces increase significantly with an increase in the number of vessel turns (Poulos et al., 2024). Similarly, another study shows that removal forces are significantly higher in smaller distal vessels than in large vessels (Tsuto et al., 2024).

Finally, the clot-SR friction coefficient plays an important role in clot entrapment during retrieval. Still, a higher friction might fragment the RBC-rich clot and cause a secondary embolism in distal intracranial circulation (El-Bouri et al., 2021; Luraghi et al., 2022; Nagargoje et al., 2025). Higher clot-SR friction might help retrieve fibrin-rich clots, but the inability of SR penetration in the fibrin-rich clots is a major issue. New SRs are being developed for penetrating stiff fibrin-rich clots using helical designs and are proving beneficial (Lehnen et al., 2023; Nordmeyer et al., 2023). Although we can find multiple in vitro studies on clot-vessel friction and SR-vessel friction investigation, there is an inconsistency in various studies and significant standard deviations. Understanding the role of friction in MT is crucial for optimizing the SRs and improving the procedure outcomes.

This review focuses on understanding the physics involved in various interactions of MT and recent progress in this field. We have explored the fundamentals of friction and its application in MT. The discussion centers on the clinical implications of friction and proposed future directions for enhancing the MT procedure. Further sections discuss individual interactions in MT and their development.

## 2. Methods

A literature search was conducted in April 2025 using the keywords "mechanical thrombectomy", "friction in mechanical thrombectomy", "clinical trials in mechanical thrombectomy", and yielded 225 research articles. The search strategy utilized electronic platforms, including PubMed, Web of Science, and Google Scholar. Only articles written in the English language were selected, and articles investigating retrieval forces/friction forces in MT were included in the analysis. Based on the filter applied and followed by Preferred Reporting Items for Systematic reviews and Meta-analysis (PRISMA) guidelines, only 09 research articles were eligible for the current study, as shown in Fig. 2. Only peer-reviewed journal articles have been considered. Conference proceedings, book chapters, thesis and dissertation, and preprints were excluded from this review. The research articles published between 2015–2025 have been included in this study. The full texts of studies dedicated to investigating friction in MT were examined, analyzed, and reported in this work.

We found several articles on MT using in vitro and in silico models, and we have excluded them based on the absence of knowledge about frictional interaction and its relation to mechanical thrombectomy outcomes. The purpose of this review article is to showcase the importance of frictional interactions during MT and the unavailability of important aspects of frictional interactions during MT. Several aspects of these interactions have been discussed, and future recommendations have been made to address those drawbacks.

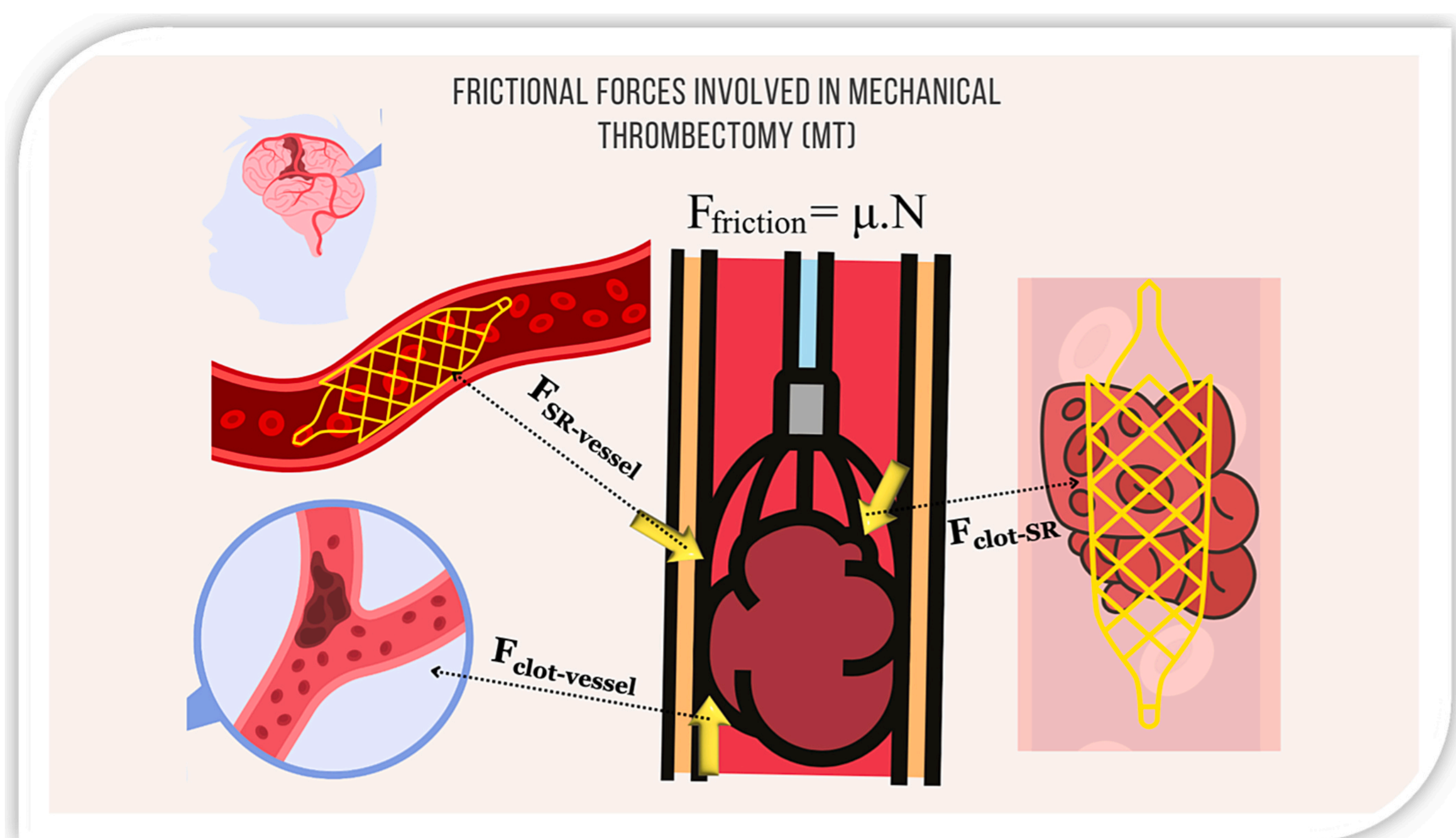

**Fig. 1.** Schematic of frictional forces involved in mechanical thrombectomy. In MT, frictional forces are developed between: i) SR and vessel, $F_{SR\text{-}vessel}$, ii) SR and clot, $F_{clot\text{-}SR}$, and iii) clot and vessel, $F_{clot\text{-}vessel}$. In the figure, N represents the normal force between the SR, clot, and vessel, generated during the interaction, whereas μ is the friction coefficient.

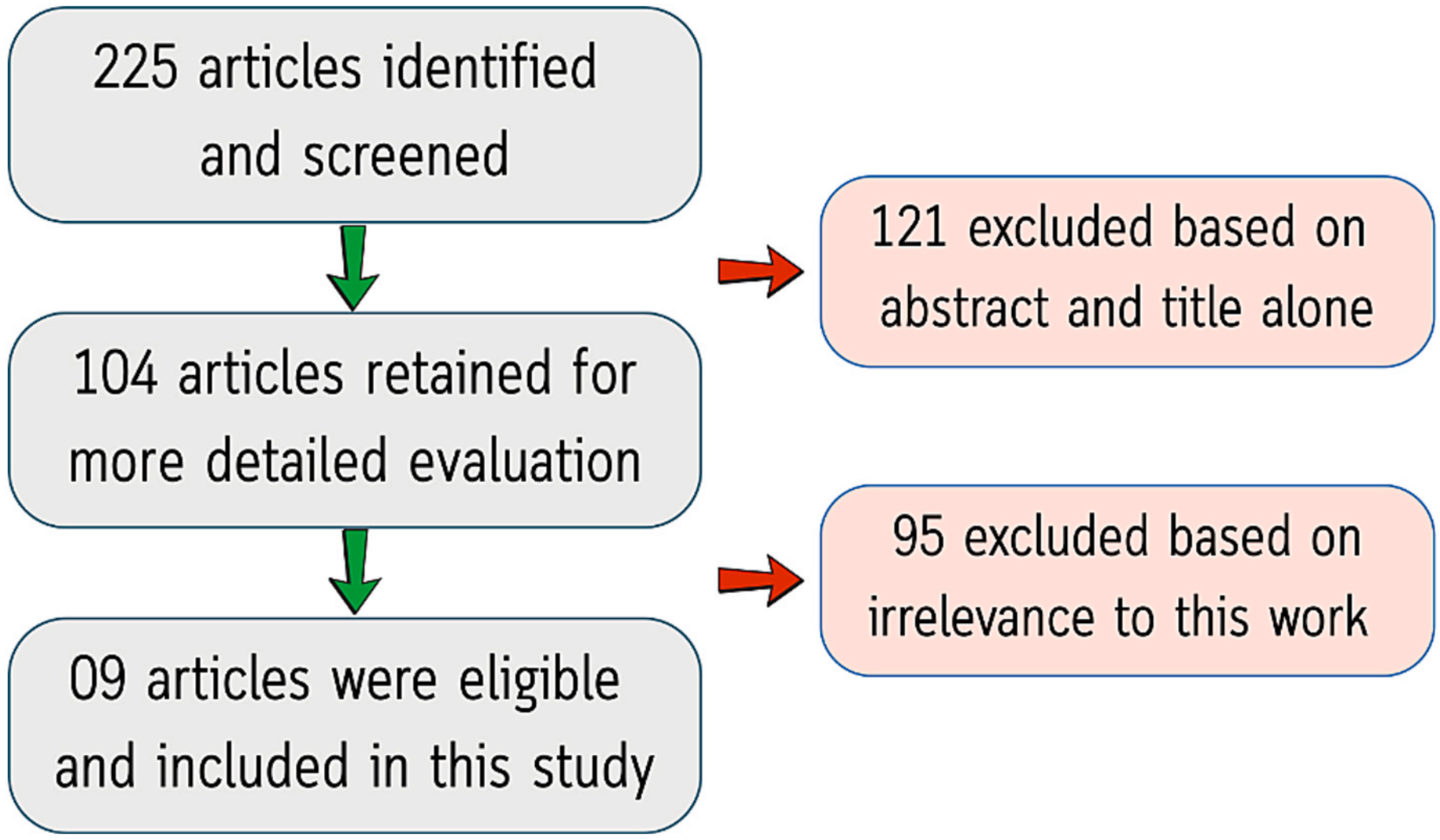

**Fig. 2.** Schematic of the PRISMA flowchart used for this study.

## 3. Mechanics of friction in MT

Friction plays a major role in removing clots from the cerebral circulation and regaining blood supply to the brain. Several attempts have been made to understand the role and importance of frictional interactions in MT outcomes. Fibrin-rich clot exhibits higher friction compared to RBC-rich clots, which aligns with clinical observations. However, the calculation friction coefficients have several limitations, such as the use of flat and artificial surfaces instead of realistic surfaces with endothelial inner layer, and thus requires further investigations. The concept of friction was first proposed and studied by Leonardo Da Vinci around 1493, and sketches of it can be found in his notebook Codex Atlanticus (Hutchings, 2016). He proposed the concept of dry friction, and these observations were rediscovered by Amontons and Coulomb and formulated into the mathematical expression $F = \mu N$ , where $\mu$ is the friction coefficient and $N$ is the normal load. The same expression has been used in the in vitro and in silico models for calculating friction coefficients between various interactions. During MT, friction forces are generated because of the interaction between clot and vessel, clot and SR, and vessel and SR, as shown in Fig. 1. In clinical practice, physicians are facing challenges during the removal of fibrin-rich thrombi, and in vitro models have found similar observations. Various researchers justified the difficulty in removing fibrin-rich thrombi due to their higher friction values. We have discussed the importance and implications of friction in separate sections: i) clot-vessel interaction, ii) clot-SR interaction, and iii) vessel-SR interaction.

### 3.1. Clot-vessel interaction

Various factors determine the forces required to retrieve the clot, as shown in Fig. 3. The primary well-studied and major factor is the friction between the clot and the vessel wall. The value of friction and MT outcome depends on the type of clot. Fibrin-rich clots have shown higher friction coefficients and are more resistant to retrieval (Gunning et al., 2018; Machi et al., 2017). The friction coefficients for clots have been measured using non-biological materials and biological tissue. Biological tissues show significantly higher friction coefficients than non-biological materials (Elkhayyat et al., 2024; Gunning et al., 2018). The difference in friction coefficient between fibrin-rich and 20 % RBC-rich clots is significant in specific biological tissue and non-biological materials. The difference is negligible between 20 %-100 % RBC-rich clot. It might be due to structural differences in clots such as porosity and permeability.

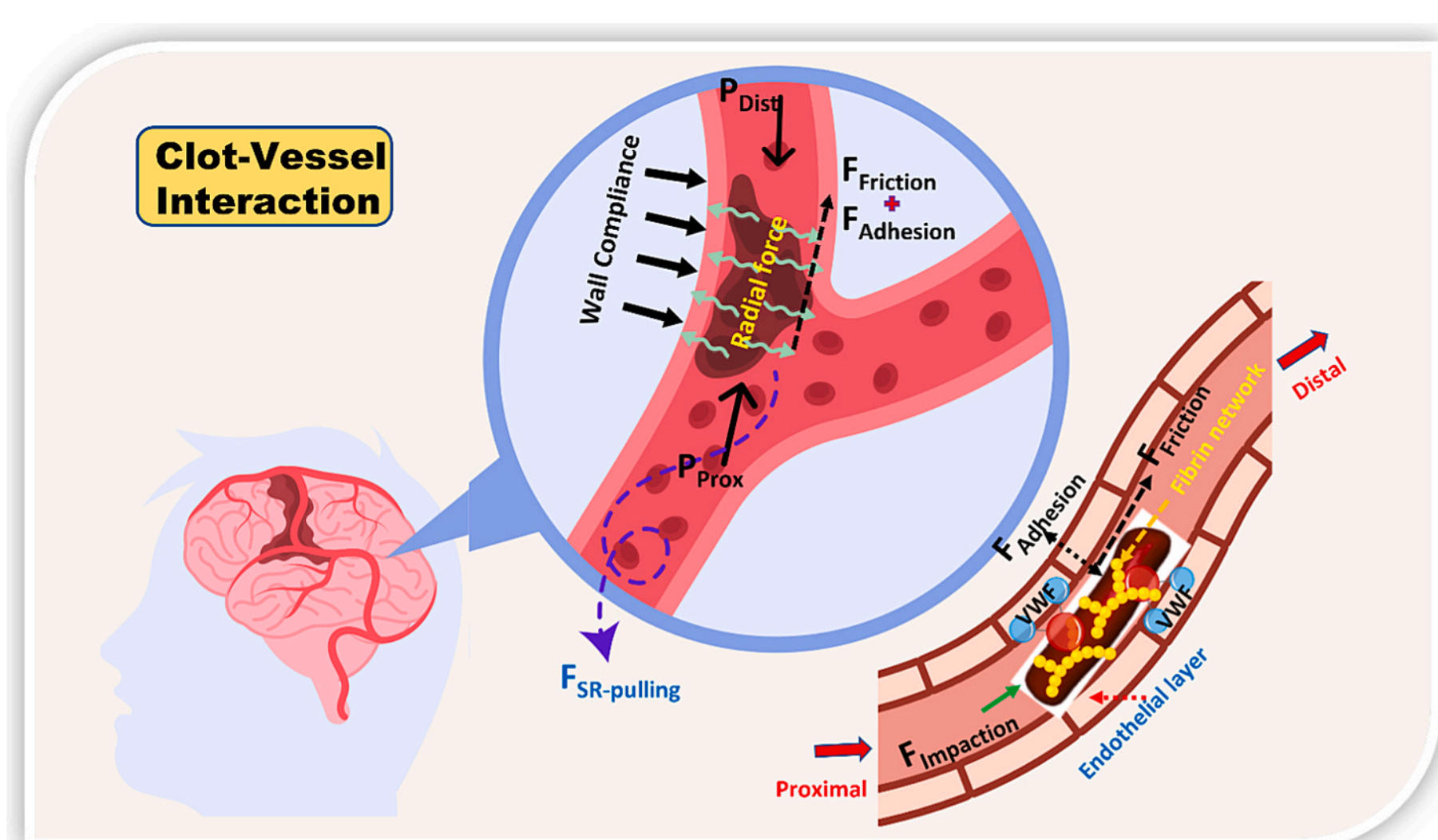

**Fig. 3.** Clot-vessel interaction and various forces involved. The blood pressure acting on the free surfaces ($P_{Dist}$ and $P_{Prox}$) establishes a pressure gradient along the clot that contracts the clot, causing a radial force against the vessel wall that generates a friction force, $F_{Friction}$, in addition to the adhesive force, $F_{Adhesion}$, between the vessel and the clot.

Porosity shows the relation of empty spaces filled with fluid in clot structure compared to the total clot volume. RBC-rich clots have more open spaces due to looser fibrin network, which leads to ease in penetration of fluids and SR. In contrast, fibrin-rich clots are densely packed and very little open space to fluid and SR penetration. RBC-rich clots have larger pores and higher permeability for fluids. While fibrin-rich clots have smaller pores and lower permeability for fluids (Carr and Hardin, 1987). Fibrin-rich clots are mechanically stiffer and less deformable. When dragged during retrieval, instead of deforming and sliding smoothly like RBC-rich clots, they maintain their shape and rubs harder against vessel wall. Due to harder rubbing against wall, fibrin-rich clots show higher friction. The rigid fibrin strands can interlock with endothelial microstructures, acting like tiny hooks. This increases the pulling force during retrieval and can cause endothelial tears. During multiple SR passes, clot contraction occurs and leads to denser clot volume. Clot contraction mechanism shows that fibrin-platelet mesh moves to the outer side of the clot surface and RBCs get packed in the center changing shapes from discs to polyhedral (Litvinov and Weisel, 2023). The clot compression increases the stiffness of clot, which leads to higher friction coefficients. Higher friction from compressed clots and vessel walls leads to endothelial cell damage, wall remodeling, and vasospasm. The retrieval behavior depends on the type of clot; RBC-rich clots show lower elastic modulus (deformable), while fibrin-rich clots show a higher elastic modulus (stiffer) (Cruts et al., 2023). However, RBC-rich clots fragment at higher friction and may cause distal secondary embolization. Fibrin-rich clots become even stiffer after interacting with the SR/aspiration catheter, making them difficult to entrap and retrieve after first-pass failure (Boodt et al., 2021).

The secondary factor responsible for the removal forces is the pressure gradient along the clot. The higher the pressure gradient, the stronger the lodging pressure and impaction force. Pressure gradient is difference in pressure across length of the clot, which drives blood flow and influence the force pushing the clot. Lodging pressure is the local pressure exerted by the clot as it lodged in vessel narrowing and decides how firm the clot gets lodged. Impaction force is the net mechanical force pushing the clot into vessel wall, arising from combination of lodging pressure and hemodynamic forces from pressure gradient. Higher impaction force increases the clot contact with vessel wall and clot gets deeply embedded into endothelial irregularities. In such cases, static friction will be much higher due to the higher impaction force. While vessel compliance reduces the normal force pressing the clot into narrowing vessels. It has been argued that the impaction force can be overcome by deploying the SR or using a combined technique with aspiration close to the clot. SR deployment enlarges the vessel due to the radial force of SR, and the clot becomes much easier to remove by overcoming the impaction force. Clinical observations found that at lower collateral pressure ($P_{dist}$) (higher pressure drop), the removal of clots becomes difficult (Maïer et al., 2020). The higher collateral pressure (lower pressure drop) pushes the clot in the opposite direction of the primary blood flow and helps the SR to overcome the static friction between the clot and vessel.

Clot lodging alters the local hemodynamics and endothelial cells become activated due to shear stress alteration or direct embolus contact. It leads to release of Von Willebrand Factor (VWF) and P-selectin, it promotes the cell and clot adhesion. Upon release, VWF forms elongated strings that act as molecular bridges, enabling blood clots to adhere to the endothelium, as shown in Fig. 3. Under high-shear conditions, this mechanosensitive glycoprotein mediates thrombosis (Zhussupbekov et al., 2022), and at extreme shear rates, VWF-dependent interactions become the dominant mechanism initiating platelet adhesion (Kim et al., 2019; Rana et al., 2019). In addition to binding platelets, VWF also binds to collagen in the subendothelial matrix, further strengthening clot adhesion. Adhesion between the vessel wall and clot is an important aspect of MT, but it has not been considered in the past in silico and in vitro studies (Kwak et al., 2022; Luraghi et al., 2021b, 2021a; Machi et al., 2017; Tsuto et al., 2024). However, a previous study demonstrated that embolus analogues adhered closely to the vessel wall through interactions involving GP IIb/IIIa and fibrin at the interface, with quantifiable adhesion forces confirming stable clot-vessel binding in a rabbit CCA stenosis model (Chueh et al., 2012). Few recent studies have investigated the adhesion strength of clot to vessel wall and cardiovascular device materials, showing that fibrin-rich clots adhere more strongly than RBC-rich clots (Good et al., 2020), micro-scale surface roughness increases adhesion (Kannojiya et al., 2024), clot age and dehydration elevate adhesion strength to synthetic polymers (Liu et al., 2024), and braided stent retrievers produce greater clot–device contact area and adhesion than laser-cut designs (Patki et al., 2023). In case of no damage to endothelial, clots consist of fibrin networks, a major structural protein, which can physically entangle and chemically bind with molecules on the endothelial surface. Endothelial damage due to compromised hemodynamics and embolus leads to exposure of collagen types I and IV, laminin, and fibronectin. These proteins promote binding to fibrin, platelets, and integrins present on the embolus surface (Kleeberg et al., 2025).

Clot composition strongly influences these interactions. Fibrin-rich clots, with their dense fibrin network, enhance adhesion both through biochemical binding and through mechanical interlocking with endothelial surface irregularities. Blood clots exhibit the highest adhesion energy when in contact with muscle tissue and the lowest with the inner arterial lining, while increased red blood cell content raises both adhesion and fracture resistance. In contrast, higher platelet content increases adhesion but reduces bulk fracture energy (Liu et al., 2024). These biomechanical and biochemical factors together can increase resistance to retrieval during mechanical thrombectomy (Duffy et al., 2019), and fibrin-rich clots, in particular, have been reported to be more difficult to retrieve (Yuki et al., 2012).

Few past studies have attempted to calculate clot-vessel friction coefficients, and some in silico studies have predicted MT outcomes using these values; however, these studies modeled only frictional interactions and not adhesion (Elkhayyat et al., 2024; Gunning et al., 2018; Johnson et al., 2017; Luraghi et al., 2021b). It might be one of the reasons for the disparity in the success ratio of thrombectomy in clinical practice and in vitro/in silico studies. In vivo, adhesion may play a major role in removal forces and needs to be considered in future studies. Adhesion (a biochemical phenomenon) is a distant from friction (a mechanical phenomenon): adhesion involves interfacial energy and bond formation, whereas friction is resistance to sliding. A recent study developed an empirical correlation based on imaging dynamic permeability for identifying clot types and adhesion mechanisms before the MT procedure (Anagnostakou et al., 2024). They found that distinguishing fibrin-rich clots from BRC-rich clots is possible during MT planning.

Another important aspect is the injury caused due to higher friction. The clot-vessel interaction causes the endothelial cell layer damage, especially from stiff fibrin-rich clot and leads to vasospasm. The adherent clot may trigger localized endothelial activation and releasing vasoactive mediators such as endothelin-1 and nitric oxide. Perivascular nerves may induce vasoconstriction around the occlusion site to minimize downstream damage, but it can affect reperfusion negatively even after clot removal (Findlay et al., 2016; Macdonald, 2016). The vasospasm due to clot occlusion is understudied and can be explored further by using bioengineered vessels to understand endothelial damage for various clot stiffness.

### 3.2. Clot-SR interaction

SRs are an integral part of MT and have emerged as a breakthrough in the AIS treatment. One of the causes of the success of MT is the optimum frictional coefficient between SR and clot. A recent in silico frictional sensitivity analysis found that low clot-SR friction shows slippage of the clot, while too high friction causes RBC-rich clot fragmentation (Nagargoje et al., 2025). The frictional interaction between clot-SR is complex and depends on various factors such as SR design, SR-clot friction, clot type, and compression/deformation of the clot, as shown in Fig. 4. Investigation of various parameters is crucial in improving the outcomes of MT by optimizing the above-mentioned factors.

The MT is performed in two stages: preliminary (SR crimping and SR tracking) and actual procedure (SR deployment and SR retrieval), which have been mimicked by our group using in silico models in past studies (Luraghi et al., 2021b, 2021a). The SR, validated through experimental stress tests, is crimped within the microcatheter and navigated along the vessel centerline to the vicinity of the occlusion site. Deployment is performed by holding the SR in position while retracting the microcatheter, allowing the SR struts to expand and capture the clot. The SR deployment position relative to the clot has been investigated and observed that the longer SR and one-third length distal to the clot resulted in better recanalization rates (Girdhar et al., 2020). The active push deployment (APD) technique has performed better than conventional simple unsheathing. The pushing of SR inside the clot causes radial bulging of clot and axial compression, which results in stiffening of clot and difficult to retrieve after first pass (Wiesmann et al., 2017).

The first pass effect (recanalization at the first attempt) strongly correlates with good long-term outcomes, and it depends on the SR design and clot properties. It has been observed using in vitro models that RBC-rich clots are easy to retrieve in comparison with fibrin-rich clots (Machi et al., 2017; Poulos et al., 2024), as shown in Fig. 4. A similar observation has been confirmed by in silico studies (Luraghi et al., 2021b). RBC-rich clots are easy to capture in the SR cage; however, lower friction between SR and clot leads to dislodgement, whereas higher friction can result in fragmentation, as observed in recent in silico frictional sensitivity analysis (Nagargoje et al., 2025). The ranges of lower and higher friction coefficients were adapted from in vitro studies, and these values fall within the reported standard deviations. In contrast, fibrin-rich clot is stiff and entrapping the clot into SR is challenging; the clot tends to roll between vessel wall and the SR struts (Machi et al., 2017). SR or aspiration devices apply a tensile force, which causes RBC-rich clots elongate and fragment (Liu et al., 2021). During retrieval, SR devices apply a tensile force on the clot as they are pulled axially. The entrapped clot in the SR is stretched until it breaks, with adhesion and frictional forces opposing retrieval. The fragmentation of the RBC-rich clot is an event that increases the procedural time and reduces the chances of good outcomes. Clot fragmentation may cause a secondary distal embolization in narrowed vessels.

On the contrary, fibrin-rich clots are stiff and difficult to entrap and retain in SR cages. The stiff clots roll between SR and the vessel wall instead of entrapment in the SR cage (Machi et al., 2017). It has been observed that aspiration catheter alone is effective for RBC-rich clots and SR with balloon-guided catheter (BCG) is better for fibrin-rich clots (Madjidyar et al., 2020). If only SR is being used for fibrin-rich clots, then longer lengths of SR are effective (Girdhar et al., 2020), and the use of a balloon guide catheter increases the chances of successful recanalization (Madjidyar et al., 2020). The mechanism of clot entrapment based on clot type remains unexplored and needs to be understood in relation to their microscopic structures after retrieval. A recent study used high-resolution scanning electron microscopy to examine entrapped clots from AIS patients and identified different modes of entrapment depending on clot composition (Dumitriu LaGrange et al., 2022). Clot capture by the SR is strongly adhesive in the case of porous and deformable clots, whereas it is weak and non-adhesive for stiff clots. As discussed in the previous section, the fibrin network, stronger in stiff clots, forms a strong molecular bond with the endothelial cell layer, increasing adhesion.

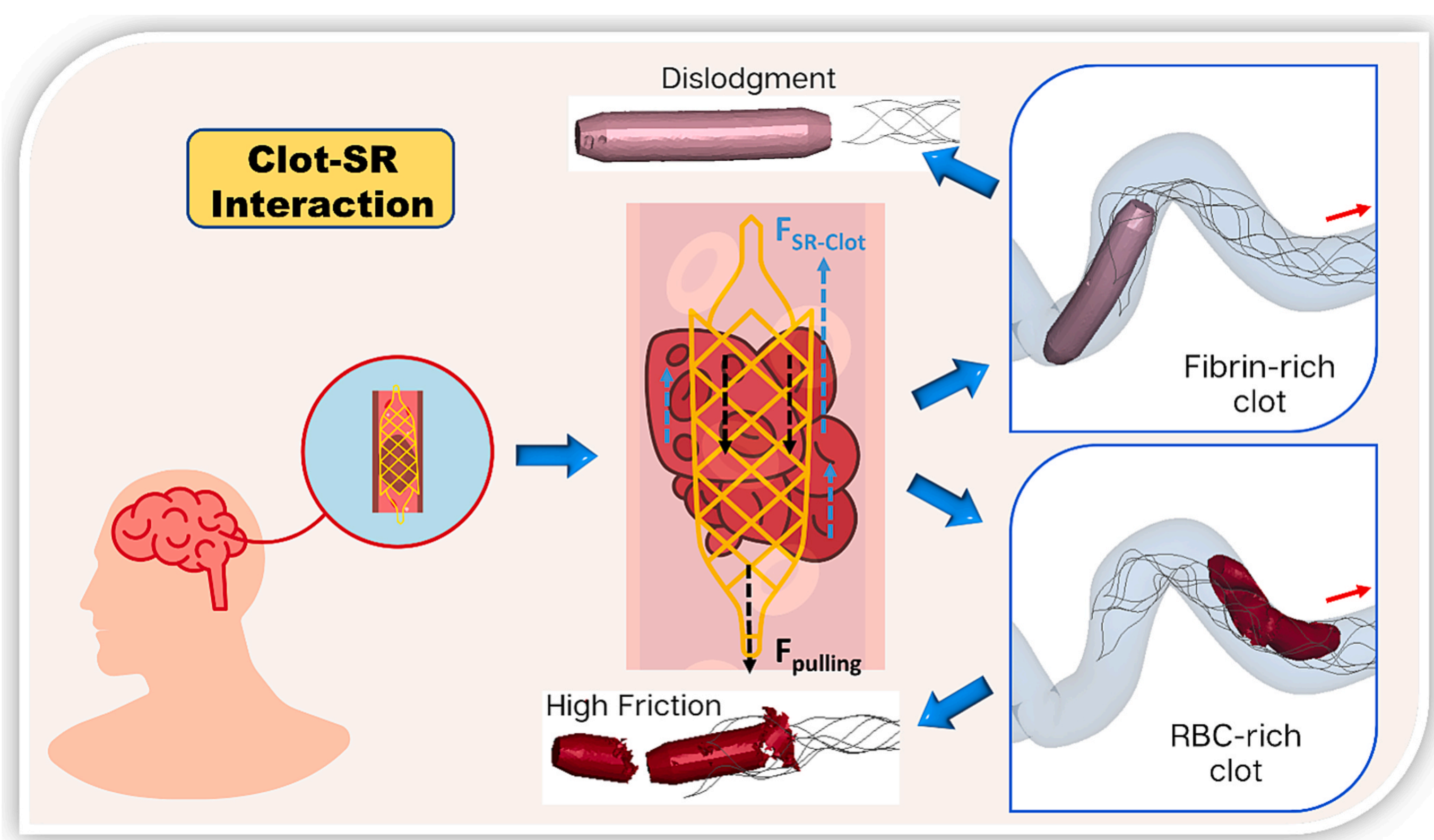

**Fig. 4.** Clot-SR interaction and clot behavior for RBC-rich and fibrin-rich clots. RBC-rich clots are easily entrapped in the SR and difficult to keep intact, which leads to distal embolization. Whereas, fibrin-rich clots tend to roll between the SR and the vessel wall, which leads to dislodgement.

Recently, new SRs have emerged to entrap the fibrin-rich stiff clots (Fennell et al., 2018; Kaneko et al., 2019; Ribo et al., 2020). The failure in the first pass leads to clot compression and becomes difficult to retrieve during subsequent passes. Balancing sufficient friction to ensure clot retention against minimizing clot fragmentation represents a critical design trade-off in SR devices. This can be conceptualized as 'contact tuning,' where surface properties and device–clot interactions are optimized to achieve the desired balance. In silico studies, such as parametric sweeps or surrogate modeling, could systematically explore these parameters and guide device designs toward optimal performance. A recent study found that SR coated with DNA-binding compounds improved clot retrieval efficiency during MT procedure and reduces the distal embolization risk (Skarbek et al., 2023). In silico studies will be a great choice for investigating such parameters and optimizing the friction between SR and clot. A previous study developed an in vitro model using radiopaque clot analogs and vascular phantoms to quantify the clot integration factor (CIF)-the amount of clot migrating into the inner volume of an expanded Trevo stent retriever-employing image segmentation techniques (Van Der Marel et al., 2016). The authors found that using a forward-push deployment technique significantly enhanced clot integration in hard clots, and that waiting approximately five minutes after deployment increased CIF across clot types, suggesting time-dependent embedment improves device–clot engagement.

### 3.3. SR-vessel interaction

The interaction between stent retrievers (SR) and vessels is another crucial factor affecting the safety and success of mechanical thrombectomy (MT). Multiple clinical studies have observed that MT outcomes are suboptimal in vessels with acute curvature, leading to lower recanalization rates and longer procedure times (Hoshino et al., 2024; Kim et al., 2022; Koge et al., 2022; Saber et al., 2024; Snelling et al., 2018). Optimum interaction achieves balanced contact that maximizes clot engagement while minimizing vessel wall injury. The friction between the SR and the vessel wall has been extensively studied, particularly regarding endothelial cell damage and its potential complications for procedural safety. A recent work quantified the endothelial cell injury from SR, and it ranged 16 %-18 % in wire/microcatheter only treatment, 37 %-61 % in 1-pass treatments, and 52 %-70 % in 2-pass treatments (Starr et al., 2025). Similarly, aspiration catheters damage the endothelial cells, especially with multiple passes (McCulloch et al., 2023). The frictional interaction arises from the interplay between the SR's radial forces and vessel compliance (Fig. 5). In curved vessel, friction also becomes a critical factor influencing clot disengagement from the SR. Notably, SR-vessel interaction plays a pivotal role in MT success, as the SR's behavior in acutely curved vessels results in compression of its cylindrical cross-sectional area-a phenomenon demonstrated in both in vivo and in vitro studies (Kaneko et al., 2019; Machi et al., 2017; Misaki et al., 2016). Due to such compression, fibrin-rich clot gets disengaged from the SR and RBC-rich clot observed fragmentation and disengagement, as shown in Fig. 5. A recent in silico study, validated using in vitro models, confirmed this SR compression and found that it increases with the friction coefficient in the curved region (Nagargoje et al., 2025). To avoid such disengagement several alternatives have been proposed: double cage structured SR-overlapping mesh layers designed to improve clot capture (Embotrap, Cerenovus, Galway, Ireland); SRs with multiple articulated segments that remain open during extension (NVI-SR, NeuroVasc Technologies, Irvine, California, USA); or SRs with multiple drop zones-regions within device with reduced radial force intended for clot accommodation and reduce fragmentation (NeVa, Vesalio, Maryland way, Brentwood, USA); that have shown better performance in curved vessels (Akpinar et al., 2021). The inner ring of these SRs gets compressed, but the outer ring covers the entire cross-section of vessels, even in highly acute curved vessels (Kaneko et al., 2019). The ThrombX retriever (ThrombX Medical Inc.) shows higher retrieval rates (90 %) in first-pass, especially in curved vessels for fibrin-rich clots (Chueh et al., 2021). On the contrary, these modified SR designs cause higher stress on the curved vessel region, which leads to endothelial cell layer damage and leads to vasospasm.

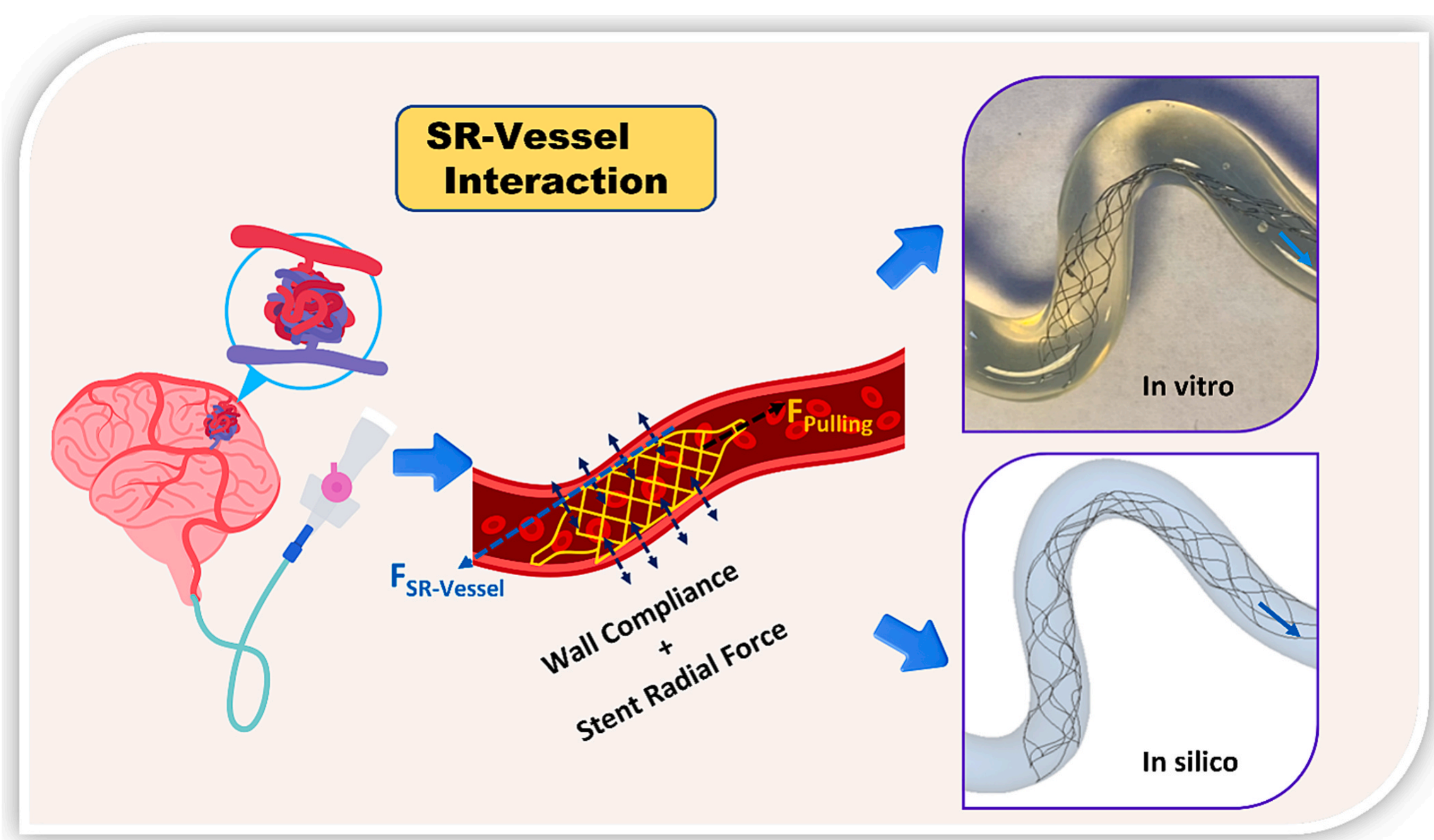

**Fig. 5.** SR-vessel interaction and its impact on curved vessels. The high friction coefficient and tortuosity causes a reduction in the effective cross-sectional area of the SR, favoring clot disengagement and MT failure.

The vasospasm occurs in 20 % of MT procedures and there are disagreements among physicians about the clinical relevance of vasospasm during MT and its management or treatment. SRs causes vasospasm and it was observed in distal regions in comparison with proximal. The frequency of vasospasm increases with increase in the number of passes of MT (Uchikawa et al., 2019). A recent study observed that vasospasm influences different treatment strategies and exists heterogeneity in concluding remarks (Jesser et al., 2024). The heterogeneity in management of vasospasm was observed due to its limited knowledge and further efforts should be focused on understanding and management of vasospasm following SR maneuvers.

It has been assumed that dynamic friction between SR and vessels is very low and has been ignored in the past in-silico models. A recent study attempted to calculate the dynamic friction values for SR-vessel interaction using tissue attached to a flat surface, whereas in vivo this interaction occurs on cylindrical, irregular surface (Elkhayyat et al., 2024). They have pulled the SR on this flat tissue and measured the friction coefficient from retrieval forces. They have analyzed using in vitro experiments that the dynamic friction coefficient between SR and vessel is significantly high, and it should be considered during in silico modeling of MT. Further efforts have been made to understand dynamic friction by quantifying the retrieval forces for various anatomies, and observed that curved vessels show significantly higher retrieval forces (Kaneko et al., 2019). While comparing SR performance between Trevo XP (older SR) and Eric (newer SR), Trevo XP showed the lowest retrieval force, and Eric 4 showed the highest (Kwak et al., 2022). The diameter of vessels played an important role in retrieval forces, and it was observed that distal smaller vessels show significantly higher retrieval forces and higher resistance to SR removal in comparison with larger vessels (Tsuto et al., 2024). The larger retrieval forces in smaller vessel might be due to larger SR diameter relative to vessel size and over constrained SR causes higher friction. Vessel diameters were more sensitive to retrieval forces than the vessel curvatures. The major issues in distal occlusion sites are vessel straightening and traction due to SR or aspiration catheter, as the distal vessels are less surrounded by subarachnoid membranes. The traction and vessel straightening leads to subarachnoid hemorrhage and rarely presents a significant complication.

In vitro studies have investigated SR-vessel friction, demonstrating its critical role in improving mechanical thrombectomy (MT) outcomes. Qualitative assessments of SR compression in acutely curved vessels exist, but quantitative data on the degree of compression and its validation through in vivo methods remain unexplored. A recent study validated the SR compression using in vitro model, but it was limited to Solitaire SR and further research is needed for various commercial SR devices compression (Nagargoje et al., 2025). Validation of such compression should be performed with in vivo by using radiopaque marker signals. Expanding this knowledge through comprehensive in vitro, in silico, and in vivo studies of various SR designs in patient-specific anatomies would significantly advance the field. Such data could help pinpoint the root causes of MT failure in tortuous vasculature by enabling detailed analysis of SR-vessel interactions.

Further research should be dedicated to a quantitative understanding of the degree of SR compression and its relation to vessel curvature, vessel diameter, and various SRs. Such studies should be extended to in vitro trials and predictive modeling of MT outcomes based on anatomic variations. The major limitations of the past in vitro studies are the use of silicon polymer for model fabrication, and they have high stiffness in comparison with in vivo vessels. It is very difficult to create or mimic in vivo vessel properties experimentally. Few research articles were able to culture endothelial cells on the inner surface of silicon models to investigate the endothelial damage caused by SR (Starr et al., 2025; Teng et al., 2015). Similar methodology could be adapted for understanding frictional coefficients and retrieval forces in realistic cerebral vessels. While in silico studies can mimic the in vivo vessel properties and show great potential for understanding SR-vessel interaction. Currently, limited knowledge is available on SR compression in acutely curved vessels, and it might contribute to the fragmentation and disengagement of clots. Many in vitro/in silico studies should be focused on understanding various commercially available SRs performances in flexible vessels and their retrieval mechanism at curved vessels. Further research should focus on understanding the performance of double-cage and multiple-drop-zone SRs and their impact on vessel injury resulting from higher retrieval forces. Clinically, it is quite well-known that the failure percentage of MT is highest in acutely curved vessels (Hoshino et al., 2024; Koge et al., 2022; Saber et al., 2024; Snelling et al., 2018). SR compression might be one of the reasons for such failures, and it has been understudied in previous studies.

Only nine studies have addressed friction, highlighting its underrepresentation in both predictive models and device evaluation, particularly in relation to dynamic friction, as shown in Table 1. Given its critical role in clot retention, device tracking, and vessel safety, friction should be more prominently integrated into in silico, in vitro, and in vivo investigations. Interdisciplinary modeling approaches-combining biomechanics, materials science, and computational simulations-could offer a more comprehensive understanding of frictional behavior and its implications for stent retriever design and clinical performance.

## 4. Challenges and future directions

Despite numerous breakthroughs and advancements in the treatment of AIS, the limited understanding of various interactions during MT hinders satisfactory long-term treatment outcomes. Lack of knowledge stems from multiple challenges in in vitro and in silico modeling, as well as its validation with clinical cases. Table 2 shows the challenges and gaps in knowledge in MT modeling and prediction of outcomes. Most in vitro MT have been conducted on generic silicon phantoms rather than patient-specific anatomies. Even when patient-derived geometries from CT scans are used, the material properties of arterial tissues are not accurately reproduced. Silicon phantoms cannot replicate the characteristics of the endothelial cell layer, and further efforts are needed to incorporate such layers into models. Because friction is a microscopic phenomenon, accurately replicating tissue properties in silicon models is essential for improving SR-tissue interaction studies. For example, performing friction coefficient measurement on a tissue layer applied over silicon phantoms could better mimic in vivo conditions. A few studies have used biological tissues to calculate static and dynamic friction, but these experiments ware conducted on flat surfaces rather than realistic vessel anatomies (Elkhayyat et al., 2024; Gunning et al., 2018). Over the biological tissues (stitched to flat surfaces), they observed significant discrepancies in friction coefficients between various studies and high standard deviation during individual studies. However, a few studies used endothelial cells layer on silicon phantom to investigate endothelial damage due to SR (Starr et al., 2025; Teng et al., 2015). A similar methodology can be extended for calculation of friction coefficients in realistic cerebral vessel silicon phantoms. It is very difficult to obtain the in vivo friction values, but using the above-mentioned suggestions, we can move slightly closer to in vivo scenarios. The use of sensor technology in calculating in vivo friction coefficients could potentially offer a novel approach, through further research and validation would be necessary to confirm its feasibility and accuracy.

Another challenge is to create blood clots that mimic in vivo properties. Most of the studies used static conditions to create blood clots, either from animal or human blood samples. Few of the recent studies used the Chandler loop dynamic condition for creating heterogenous clots from bovine blood (Madjidyar et al., 2020; Poulos et al., 2024). The issue of clot preparation is critical, as the comparison between clots formed under static and dynamic conditions highlights important mechanical differences. Clots generated under dynamic protocols-such as in a Chandler loop-exhibit denser fibrin networks, smaller pore structures, and greater stiffness compared to static clots (Eyisoylu et al., 2024; Liu et al., 2023). Incorporating these dynamics better reflects clot heterogeneity and improves the physiological relevance of in vitro thrombectomy models-strengthening predictive validity when evaluating device performance and clot–device interaction.

**Table 1**
In vitro studies dedicated to understanding friction during MT.

| Study | Retrieval SR type | Clot Type | Vessel model and material | Friction Measurement | Key Findings |
|---|---|---|---|---|---|
| (Tsuto et al., 2024) | SR (Trevo NXT) | No clot (only SR) | Ideal model (silicon) | Retrieval forces (RF) | Narrower and acutely curved vessels exhibited higher frictional forces. |
| (Reymond et al., 2024) | SR + aspiration (Solitaire) | Red & white (guar hum & borax) | Realistic model (flexible silicon) | Retrieval forces (RF) (during tracking & retrieval) | Differentiating soft and hard clots can be possible using tracking and retrieval forces at clot locations. |
| (Poulos et al., 2024) | SR (Solitaire & EmboTrap | Red & while (Bovine blood) | Realistic model (flexible silicon) | Retrieval forces (RF) | SR removal forces increased with model tortuosity. The type of clot did not influence removal forces. |
| (Elkhayyat et al., 2024) | SR (Solitaire) | Red and white (Bovine blood) | Glass, PVC, silicon, & arterial tissue | Friction coefficients (FC) | Non-biologic materials show significantly lower FCs than biological tissues. SR-vessel dynamic FC is significantly high and its value using silicon is close to in vivo. |
| (Kwak et al., 2022) | SR (three types) | No clot (only SR) | Patient-based (flexible silicon) | Retrieval forces (RF) | Trevo XP showed the lowest frictional force and Eric 4 showed the highest during retrieval. |
| (Kaneko et al., 2019) | SR + aspiration (four types) | Whole porcine blood | Realistic model (flexible silicon) | No friction measurement, captured MT images | Acutely curved vessels decrease success rate and new SR type (double caged) improves success in such cases. |
| (Gunning et al., 2018) | Flat surfaces | Red & white | PTFE & arterial surface | Friction coefficients (FC) | Fibrin-rich (white) clots have a significantly higher FC than RBC-rich clots. Arterial tissue shows higher FC than artificial surfaces. |
| (Larsen et al., 2018) | SR & prototypes | Whole blood thrombi | Curved silicon vessel | Retrieval forces (RF) | Complete wall-SR apposition shows the highest first-pass success and lowest embolism rate, but it also has the highest retraction force, which leads to endothelial cell layer damage. |
| (Machi et al., 2017) | SR (ten types) | Red & stiff white | Ideal (3D printed rigid) | Retrieval forces (RF) | White small thrombi are not entrapped in SR and Red clots are easily trapped, but more tendency to fragment. |

**Table 2**
Challenges and gaps in knowledge for clot-vessel-SR interaction.

| Area | Key Challenge | Gaps in Knowledge |
|---|---|---|
| Frictional Interaction | Limited quantification of *dynamic* friction under physiological conditions | Most studies use static or flat-surface models; in vivo curved, irregular vessel surfaces are underrepresented |
| Clot Capture Mechanism | Variable SR performance depending on clot type and structure | Microscopic mechanisms of clot entrapment by SR remain unclear |
| SR Design Parameters | Lack of systematic evaluation of design features (e.g., double cage, drop zones) | Limited evidence on how these features influence clot retention, fragmentation, and vessel injury |
| SR-Vessel Interaction | Device compression and altered contact mechanics in tortuous vessels | Poorly quantified effect on retrieval forces and clot disengagement |
| Predictive Modeling | Friction and deformation are underrepresented in computational thrombectomy models | Lack of validated multi-physics simulations incorporating friction, clot mechanics, and vessel compliance |
| Interdisciplinary Approach | Fragmented efforts across biomechanics, materials science, and clinical studies | Limited cross-disciplinary collaboration slows translation into device innovation |

In the in-silico models, despite achieving significant advancements, there remain accountable limitations. The limitation of using dry friction in in *silico* models is well-recognized, as it oversimplifies the complex adhesion phenomena observed during SR-clot-vessel interaction. This can be overcome by cohesive or adhesive interaction formulations that allow for partial bonding between contacting surfaces. Such behavior can be represented using cohesive zone modeling, where traction–separation laws define the initiation and evolution of adhesion, or through modified contact algorithms that incorporate both frictional and adhesive components. Incorporating these formulations is particularly relevant for bridging the gap between in vitro experimental findings-such as clot integration and retention-and the underlying mechanical concepts, ultimately improving the physiological relevance of computational thrombectomy models.

The anatomic models used were either realistic or patient-specific, but all the in-silico models assumed rigid vessel walls. Vessel wall compliance should be included together with the mechanical and structural properties of catheters validated from experiments. The assumption of rigid vessel walls is a common simplification in in *silico* thrombectomy models; however, it neglects the mechanical influence of vessel compliance on SR–clot–vessel interaction. Vessel compliance can be incorporated using fluid–structure interaction (FSI) approaches that couple blood flow with vessel wall deformation, or by implementing hyperelastic material models—such as Mooney–Rivlin or Ogden formulations—to capture the nonlinear elastic behavior of arterial tissue. Accounting for compliance is important because it can buffer retrieval-induced stresses, modulate the contact area between the SR and vessel wall, and influence clot deformation and retention. Including these effects could improve the predictive accuracy of computational models and their translation to clinical performance.

The stress–strain data of human clots have been used in recent studies, but the shape and structure of clots used were simplified. One of the major limitations is the validation and comparison of in silico models with in vivo MT outcomes. The importance of validation is well recognized in advancing in silico thrombectomy models toward clinical relevance. Potential in vivo metrics that could serve as validation targets include retrieval forces measured during MT procedures, first-pass and overall recanalization success rates, and the incidence or extent of clot fragmentation as observed via angiography or histological analysis. Aligning computational outputs with these measurable clinical or experimental benchmarks would strengthen model credibility and support their adoption for device evaluation and optimization. Large datasets of such simulations will be a great guide in understanding the complex nature of the MT procedure. The integration of machine learning with CT imaging offers a promising approach for pre-procedural clot characterization. Specifying which imaging-derived features could be leveraged—such as radiodensity, clot length, location, perviousness, or vessel curvature—would clarify the potential workflow and its predictive scope. There is precedent in the literature for using such features to infer clot composition and mechanical properties, with studies demonstrating correlations between radiodensity and fibrin or red blood cell content, as well as between perviousness and likelihood of successful retrieval (Boodt et al., 2021; Cruts et al., 2023). Explicitly linking these parameters to clot type could guide device selection and strategy optimization in mechanical thrombectomy. The inclusion of sensor technologies in the catheter or in the micro guidewire might help us to find the in vivo friction coefficients, and it can be useful during in silico modeling. A few studies used such sensors technologies for measuring friction and forces during navigation and it can be adapted for MT modeling (Takashima et al., 2019; Zhang et al., 2021).

## CRediT authorship contribution statement

**Mahesh S. Nagargoje:** Writing – review & editing, Writing – original draft, Visualization, Software, Methodology, Investigation, Funding acquisition, Formal analysis, Data curation, Conceptualization. **Virginia Fregona:** Writing – review & editing, Software, Resources, Methodology, Formal analysis, Data curation. **Giulia Luraghi:** Writing – review & editing, Supervision, Project administration, Methodology, Investigation, Conceptualization. **Francesco Migliavacca:** Writing – review & editing, Supervision, Project administration, Methodology, Investigation, Conceptualization. **Guglielmo Pero:** Writing – review & editing, Resources, Methodology, Investigation, Formal analysis. **Jose Felix Rodriguez Matas:** Writing – review & editing, Supervision, Resources, Project administration, Methodology, Investigation, Funding acquisition, Formal analysis, Conceptualization.

## Funding

This project has received funding from the European Union's Horizon 2020 research and innovation program under the Marie Sklodowska-Curie grant agreement No. 101104493.

FM, GL, and JFRM are partially supported by the European Union's Horizon Europe research and innovation program, grant number 101136438.

## Declaration of competing interest

The authors declare that they have no known competing financial interests or personal relationships that could have appeared to influence the work reported in this paper.

## Acknowledgments

None.